\documentclass{article}
\usepackage{epsfig}
\usepackage{graphicx}
\usepackage{amsmath,amssymb}
\usepackage{color}
\newcommand{\beq}{\begin{equation}}
\newcommand{\eeq}{\end{equation}}
\newcommand{\beeq}{\begin{eqnarray}}
\newcommand{\eeeq}{\end{eqnarray}}

\newcommand{\dd}{\partial}
\newcommand{\de}{\delta}
\newcommand{\m}{\mu}
\newcommand{\n}{\nu}
\newcommand{\ls}{\left(}

\newcommand{\rs}{\right)}

\newcommand{\al}{\alpha}
\newcommand{\be}{\beta}

\newcommand{\te}{\theta}

\newcommand{\la}{\lambda}
\newcommand{\dz}{\zeta}

\newcommand{\disn}[2]{$$\displaylines{\refstepcounter{equation}%
            \label{#1}\hskip 1em minus 1em #2\hfilneg}$$}
\newcommand{\nom}{\hfil\hskip 1em minus 1em (\theequation)}

\newcommand{\ns}{\hfill\cr\hfill}
\renewcommand{\le}{\leqslant}
\renewcommand{\ge}{\geqslant}

\begin{document}

\begin{center}\bf
HADRON AMPLITUDES IN COMPOSITE SUPERSTRING MODEL
\bigskip

V.A.Kudryavtsev, A.N.Semenova\\
Petersburg Nuclear Physics Institute \\

\bigskip

Abstract\\

Hadron vertices for $u$, $d$, $s$ quark flavours are formulated in
terms of interacting composite strings. Four-point amplitudes for
interaction of $\pi$, $K $-mesons are presented. \vspace{1pc}
\end{center}

%%%%%%%%%%%%%%%%%%%%%%%%%%%%%
\section{History}
An essential interest in string description of hadron interactions has arised as far back as forty years ago (the Nambu string \cite{Nam} and dual resonance models initiated by Veneziano's work \cite{Ven}) due to the remarkable universal linearity of Regge trajectories $\al(t)$ for meson and baryon resonances \cite{LovSh} and \cite{AnAn}. ($J=\al(M^2)=\al(0)+\al'_HM^2\, (\al'_H\approx 0,85 GeV^{-2})$, where $J, M$ are spin and mass of a resonance.) Now we have these trajectories up to $J=5$ and states for not only leading (n=0) but for the second (n=1), for the third (n=2) and even for the fourth (n=3) daughter trajectories $J_n=\al(0)-n+\al'_HM^2$, n=0,1,2,3... . See \cite{AnAn}.

However attempts to build the string model for hadron interactions have not been successful because consistent (compatible with unitarity) string models \cite{Lov}, \cite{Gr} have required intercept of leading meson trajectory $\al(0)$ to be equal to one. A shift of this value from one has led to contradiction with unitarity since in this case states with negative  norms will appear. The leading $\rho$-meson trajectory has the intercept approximately equal to one half. Just this reason led to superstring models for non-hadrons: for massless gluons (open strings) on the trajectory $J=1+\al'_PM^2$ and for massless gravitons (closed strings) on the trajectory $J=2+1/2\al'_PM^2$.

A generalization of classical multi-reggeon (multi-string) vertices \cite{Al} was suggested by one of authors in 1993 \cite{Ku} as a new solution of duality equations for many-string vertices in \cite{Al}. These string amplitudes have been used for description of many $\pi$-mesons interaction  \cite{Ku}. New string vertices \cite{Ku1} give a new geometric picture for string interaction which has a natural description in terms of  three two-dimensional surfaces (composite string) for moving open string instead of usual one surface. Additional edging two-dimensional surfaces carry quark quantum numbers (momentum, flavour, spin). This composite string construction reminds two other similar objects: a gluon string with two point-like quarks at its ends or simplest case of a string ending at two branes when they are themselves some strings. Let us note that we have no supersymmetry in the Minkovsky (target) space for this model. As before there is conformal supersymmetry (super Virasoro symmetry) on two-dimensional world surfaces. The topology of interacting composite strings allows to solve the problem of the intercept $\al(0)$ for leading trajectory and to obtain consistent string amplitudes which do not break down unitarity.

%%%%%%%%%%%%%%%%%%%%%%%
\section{General formulation of composite superstring model.}
Many-string vertices of interacting composite strings are natural generalization of corresponding many-string (many-reggeon) vertices for classical string models \cite{Al}. The $N$-string amplitude is represented by some integral over $z_i$ variables of the vacuum expectation value for a product of a vertex operator and wave functions of string states:

\disn{14}{
A_N= \int \prod_j dz_j \langle 0|V_N \prod_i \hat \Psi^{(i)}|0^{(i)}\rangle\,
\nom}
where $\hat \Psi^{(i)}|0^{(i)}\rangle\ $ is a wave function of $i$-th string state.

The $N$-string vertex operator $V_N^{NS}$ for the Neveu-Schwarz model is given by the following exponent:

\disn{19}{
V_N^{NS}=\exp \ls \frac{1}{2}\sum_{n,m,p, i\ne k}\frac{a_n^{(i)}}{\sqrt n}(U_{\varepsilon}^{(i)})_{nm}(V_{\varepsilon}^{(k)})_{mp}\frac{a_p^{(k)}}{\sqrt p} + \frac{1}{2} \sum_{n,m,p, i\ne k} b^{(i)}_{n+1/2}(U_{1/2}^{(i)})_{nm}(V_{1/2}^{(k)})_{mp}b^{(k)}_{n+1/2} \rs,
\nom}
where $(U_{j}^{(i)})_{nm}$, $(V_{j}^{(k)})_{mp}$ are the definite infinite matrices and the value of index $j$ answers to representation of these matrices according to certain conformal spin $j$ of fields: $j=\varepsilon\to0$ for the first item and $j=\frac{1}{2}$ for the second one \cite{Cor}. The upper indices $i,k=1...N$ number interacting strings. These vertices $V_N^{NS}$ have the necessary factorization and conformal properties and satisfy duality property.

It turns out there is another  operator $W_N$ having these properties and also satisfying duality property:

\disn{20}{
W_N=\sum_{n,m,k} \tilde \Psi^{(1)}_{n+\frac{1}{2}}(U_{1/2}^{(1)})_{nm}(V_{1/2}^{(2)})_{mk}\Psi^{(2)}_{k+\frac{1}{2}} \sum_{l,p,s}\tilde \Psi^{(2)}_{l+\frac{1}{2}}(U_{1/2}^{(2)})_{lp}(V_{1/2}^{(3)})_{ps}\Psi^{(3)}_{s+\frac{1}{2}} \times \dots \cr \dots \times \sum_{t,r,f}\Psi^{(N)}_{t+\frac{1}{2}}(U_{1/2}^{(N)})_{tr}(V_{1/2}^{(1)})_{rf}\Psi^{(1)}_{f+\frac{1}{2}} \equiv \cr \equiv \prod_{i=1}^N \sum_{n,m,p}\tilde \Psi^{(i)}_{n+\frac{1}{2}}(U_{1/2}^{(i)})_{nm}(V_{1/2}^{(i+1)})_{mp}\Psi^{(i+1)}_{p+\frac{1}{2}}
\nom}
where $i$, $j$ number strings in the same way as in (\ref{19}). $W_N$ is some cyclic symmetrical trace-like operator built of Fourier components of two-dimensional fields $\Psi ^{(i)}$. The generalized $N$-string vertex operator for composite strings is the product of the old operator $V_N^{NS}$ and the new one $W_N$:

\disn{21}{
V^{comp}_N=W_NV_N^{NS}
\nom}
It is obvious that operators $V_N^{NS}$ and $W_N$ have different structure. Matrices $U^{(i)}$ and $V^{(j)}$ in $V_N^{NS}$ connect all possible fields $a_n^{(i)}$ with each other. However, in the operator $W_N$ matrices $U^{(i)}$ and $V^{(j)}$ connect only neighbour fields $\Psi^{(i)}$ and $\Psi^{(i+1)}$.

\begin{figure}[h!]
\center{\includegraphics[width = 8 cm]{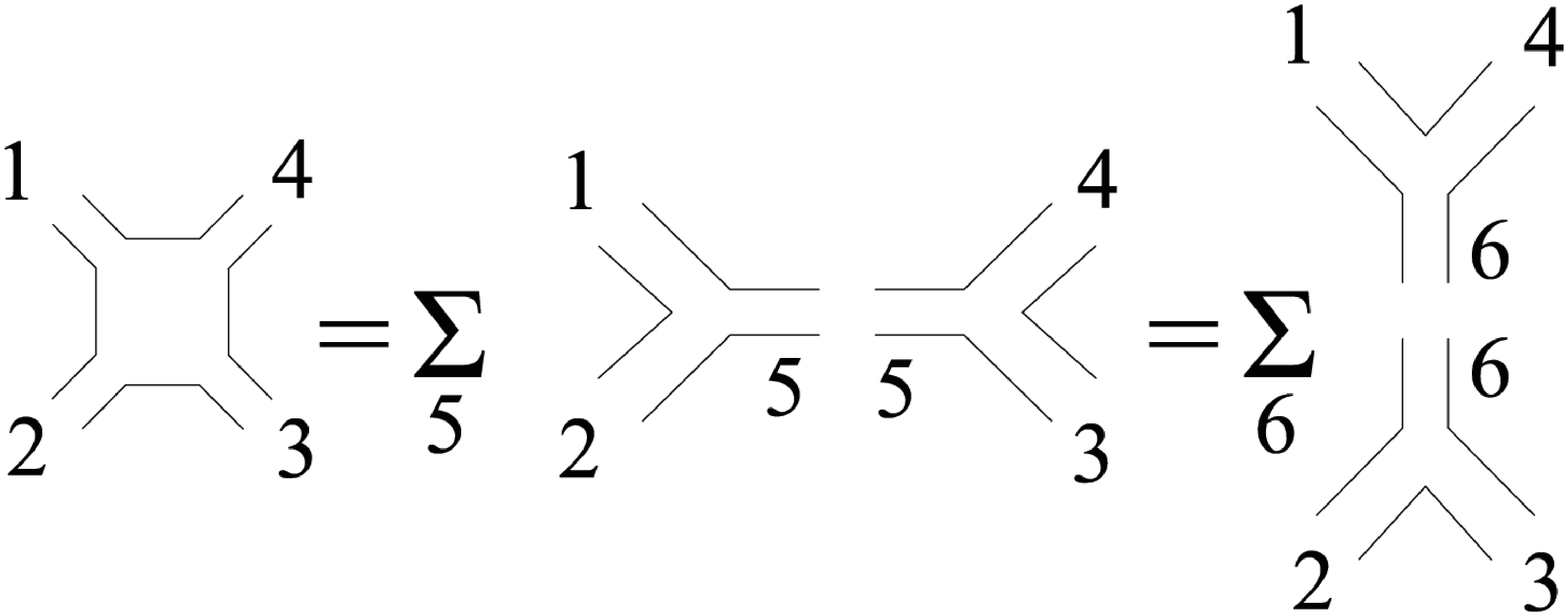}
\caption{Duality property.}
\label{dual}}
\end{figure}

That is why this operator reproduces the structure of dual quark diagrams. These diagrams display the well known OZI rule. So operator (\ref{21}) can be interpreted as vertex operator of composite strings interaction. It must be stressed that necessary factorization and duality property is satisfied in (\ref{21}), as shown in Fig.~\ref{dual}. Duality property automatically leads to the satisfaction of one-particle unitarity in $s$- and $t$-channel, if physical states spectrum is free of negative norm states.

\begin{figure}[h!]
\center{\includegraphics[width = 7 cm]{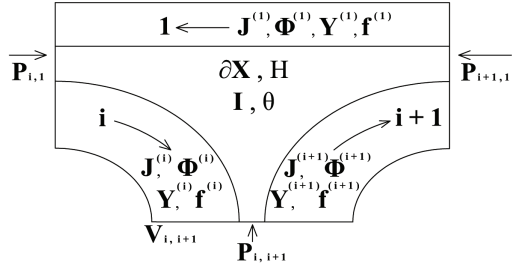}
\caption{Topology of new vertex and location of the fields on the surfaces.}
\label{vershina}}
\end{figure}

For investigation of composite superstrings it is more convenient to move from multi-string vertices to more simple case corresponding to the emission of the ground state. In this case composite superstrings vertex operator will contain additional (to usual $\dd X_{\m}$ and $H_{\m}$ fields on the basic two-dimensional surface) fields on the edging surfaces: $Y_{\m}$ and its superpartner $f_{\m}$ with Lorentz indices $\m=0,1,2,3$. We also include other fields (additional "fifth" components of fields) $J$ and its superpartner $\Phi$ which carry internal quantum numbers on edging surfaces and similar $I$ and $\te$ fields on the basic two-dimensional surface (See Fig.~\ref{vershina}).

Since the edging fields are propagating only on the own edging surface between neighbouring vertices we will have to number these fields in accordance with numbers of these edging surfaces. This is similar to numbering of strings in many-string vertices (\ref{19}). For the ground state emission vertex (Fig.~\ref{vershina}) the vertex operator $\hat V_{i,i+1}$ will contain the fields of $i$-th and $(i+1)$-th edging surfaces.  The following description of interaction amplitudes of ground string states is similar to (\ref{21}):

\disn{22}{
A_N=\int\prod dz_i\langle0|\hat V_1(z_1)\hat V_2(z_2)\dots\hat V_{N-1}(z_{N-1})\hat V_N(z_N)|0\rangle, \cr \hat V_i(z_i)=z_i^{-L_0}V_i(1)z_i^{L_0}, \cr \hat V_i(1)=g(\tilde\Psi_i(1)\Gamma_i\Psi_{i+1}(1)e^{ik_iX(1)}),
\nom}
where $\Psi_i$ is the field propagating on the $i$-th edging surface.

However, we can give another equivalent form of vertex operator $\hat V_i(1)$  in (\ref{22}) without numbers. For one set of $\Psi$ fields it has the following form:

\disn{33}{
\hat V_i(1)=g(\tilde\Psi(1)\Gamma_{(i)}|0^{\Psi}\rangle\ \langle0^{\Psi}|\Psi(1)e^{ik_iX(1)}),\cr \hat V_1(1)=\tilde\Psi(1)\Gamma_{(1)}\Psi(1)e^{ik_1X(1)}, \cr \hat V_N(1)=\tilde\Psi(1)\Gamma_{(N)}\Psi(1)e^{ik_NX(1)}.
\nom}
The operator $|0^{\Psi}\rangle\ \langle0^{\Psi}|$ allows fields $\Psi$ to be propagating between neighbouring vertices only. Due to these operators this form (\ref{33}) excludes composite string model from the set of additive string models of the Lovelace's paper \cite{Lov} and leads to the topology of composite string model (Fig.~\ref{vershina}).

It is convenient for discussion of spectrum of states to use the equivalent formulation of amplitude $A_N$. We place vacuum states in another way: outside of the vertex operator. In this case we have to keep numbers of edging surfaces. So for the middle vertices we have (for $1, 2, N-1, N$ vertices it will have some evident changes):

\disn{34}{
\hat V_{i,i+1}(z_i) \longrightarrow |0^{(i-1)}\rangle\ \hat V_{i,i+1}(z_i)\langle0^{(i+2)}|
\nom}

Now we can write the expression for amplitude:

\disn{1}{
A_N=\int \prod dz_i \langle 0^{(1,2)}|\hat V_{12}(z_1)\langle0^{(3)}|\hat V_{23}(z_2)\langle0^{(4)}|\hat V_{34}(z_3)|0^{(2)}\rangle\ \langle0^{(5)}| ... |0^{(i-1)} \rangle\ \times \ns \hat V_{i,i+1}(z_i)\langle0^{(i+2)}| ... |0^{(N-2)}\rangle\  V_{N-1,N}(z_{N-1})|0^{(N-1)} \rangle\ V_{N,1}(z_{N})|0^{(N,1)} \rangle\
\nom}
This form allows us to see symmetries of composite string model and this model to be going beyond the framework of additive string models.

%%%%%%%%%%%%%%%%%%%%%%%%
\section{Symmetries}
\begin{figure}[h!]
\center{\includegraphics[width = 8 cm]{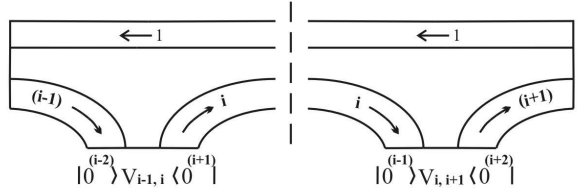}
\caption{$i$-th section.}
\label{2versh}}
\end{figure}

Main symmetry of any string model is the superconformal symmetry to be defined by the Virasoro operators $G_r$. For composite superstring model we consider the set of states and of superconformal generators for the $i$-th section between the $\hat V_{i-1,i}$ vertex and  $\hat V_{i,i+1}$ vertex in (\ref{1}). (See Fig.~\ref{2versh}.) Namely we have fields on $i-1$, $i$, $i+1$ edging surfaces: $(Y^{(i-1)},f^{(i-1)})$; $(J^{(i-1)},\Phi^{(i-1)})$; $(Y^{(i)},f^{(i)})$; $(J^{(i)},\Phi^{(i)})$; $(Y^{(i+1)},f^{(i+1)})$; $(J^{(i+1)},\Phi^{(i+1)})$ fields in addition to fields on the basic surface: $(\dd X,H)$; $(I,\Theta)$.

In order to consider our spectrum of states in more symmetric way  in regard to left and right sides we introduce a set of auxiliary fields $((Y^{(a)},f^{(a)}))$;$(J^{(a)},\Phi^{(a)})$ instead of $(i-1)$-th and $(i+1)$-th in decomposition of unity in the $i$-th section:

\disn{26}{
1=\sum|State(i-1,i,i+1)\rangle\ \langle State(i-1,i,i+1)|
\nom}

Taking into account the operator $ \langle 0^{(i+1)}|$ from the left side and operator $|0^{(i-1)}\rangle\ $ from the right side we can replace (\ref{26}) by the following sum:

\disn{27}{
\sum|State(i-1,i)\rangle\ \langle State(i,i+1)|= \cr =\de(fields(i-1)-fields(a))\sum|State(a,i)\rangle\ \langle State(i,a)|\de(fields(a)-fields(i+1))
\nom}

Here we mean for $\de(fields(i-1)-fields(a))$ (in case of $Y$-fields):

\disn{28}{
\de(Y(i-1)-Y(a))=\sum_{[\la_1,\la_2,...]}\prod_1^n\frac{(Y_{-n}^{(i-1)})^{\la_n}}{\sqrt{\la_n!}}|0^{(i-1)}\rangle\ \langle 0^{(a)}|\prod_1^n\frac{(Y_{n}^{(a)})^{\la_n}}{\sqrt{\la_n!}}
\nom}
and for $\de(fields(a)-fields(i+1))$ correspondingly:

\disn{29}{
\de(Y(a)-Y(i+1))=\sum_{[\la_1,\la_2,...]}\prod_1^n\frac{(Y_{-n}^{(a)})^{\la_n}}{\sqrt{\la_n!}}|0^{(a)}\rangle\ \langle 0^{(i+1)}|\prod_1^n\frac{(Y_{n}^{(i+1)})^{\la_n}}{\sqrt{\la_n!}}
\nom}

It is worth to note that the sets of states on the left and on the right of the vertex $\hat V_{i,i+1}$ will be the same ones that allows to connect corresponding decompositions unambiguously and to see independence on the number of the section for this consideration.

We shall consider this operator $\sum|State(a,i)\rangle\ \langle State(i,a)|$ which represents unity operator in the Fock space for $|State(a,i)\rangle\ $ and we shall extract spurious states in order to find the physical states spectrum.

Now superconformal generators $G_r$ can be defined as the following ones:

\disn{30}{
G_r=G_r^{Lor}+ G_r^{Int}
\nom}

\disn{31}{
G_r^{Lor}=\frac{1}{2\pi}\int_0^{2\pi}d\tau[(H^{\m}\frac{d}{d\tau}X_{\m}+\hat P_{\n}H^{\n})+(Y_{\m}^{(a)}f^{(a)\m}+\hat p_1f^{(1)}+Y_{\m}^{(i)}f^{(i)\m})]e^{-ir\tau}
\nom}

\disn{32}{
G_r^{Int}=\frac{1}{2\pi}\int_0^{2\pi}d\tau[(I\Theta+\xi_1\Phi^{(1)})+(J^{(a)}\Phi^{(a)}+J^{(i)}\Phi^{(i)})]e^{-ir\tau}
\nom}

Here we have $a=i-1$ (on the left side) or $a=i+1$ (on the right side). Taking into account the expressions (\ref{30}) - (\ref{32}) we can derive the corresponding commutation relations for $G_r$:

For example
\disn{44}{
\{G_r,G_s\}\ =\ 2L_{r+s}+\frac D2\left(r^2-\frac14\right)\delta_{r,-s}
\nom}

\disn{45}{
D=3d^{Lor}+3d^{Int}
\nom}
here $d^{Lor}$ is the number of $Y$(or $X$)-components, $d^{Int}$ is the number of  $J$(or $I$)-components.

\disn{46}{
 [L_n,G_r]= (\frac{n}{2}-r)G_{n+r}.
\nom}

 Let us notice that our construction of $\pi$- meson contains some definite combinations of fields with Lorentz indices:

\disn{48}{
k_i\widetilde f^{(i)}= k_i (f^{(i)} + \hat{\beta}_{i} H);\cr
k_i \widetilde Y^{(i)}= k_i (Y^{(i)}+ \hat{\beta}_i \partial X )
\nom}
and with internal quantum numbers:

\disn{49}{
\zeta_i \widetilde \Phi^{(i)}=\zeta_i(\Phi^{(i)}+\hat{\alpha}_i\Theta);\cr
 \zeta_i\widetilde J^{(i)}= \zeta_i(J^{(i)}+\hat {\alpha}_i I)
\nom}

Let us consider the construction of the spectrum generating algebra for this composite superstring in similar way as in classical string models \cite{Gr}. For the given $i$-th section (between $V_{i-1,i}$ and $V_{i,i+1}$) we have fields on $(i-1), i, (i+1)$ edging surfaces and fields on the basic surface. Spurious states for this basis are defined by products of operators $G_r$ and $L_n$. But only these states are not able to save from negative norms the spectrum of physical states as it has taken place for usual classical string models since the capacity of those of them which have negative norms is not enough. For the Fock space under consideration we can obtain states with negative norms not only as the powers of time components of the $\partial X $ and $H $ fields on the basic surface but as odd powers of other time-like components: $k_{a}\widetilde f^{(a)},k_{i}\widetilde
f^{(i)}$ and $k_{a} \widetilde Y^{(a)},k_i \widetilde Y^{(i)}$.

Additional conditions for the composite string model are required in order to eliminate all negative norms from the spectrum of physical states. There is a simple solution for it. We shall require as gauge conditions the supercurrent conditions generated  by   $k_i \widetilde {f}^{(i)}$.

Namely we shall take the following constraints for our vertices:

\disn{50}{
[k_i \widetilde Y^{(i)}_n,\hat{W}_{i,i+1}] =[\hat{W}_{i,i+1},k_{i+1} \widetilde Y^{(i+1)}_n]=0
\nom}
Then we shall have enough states of negative norms generated by all gauge constraints. The equations (\ref{50}) lead  to the conditions:

\disn{51}{
k_i^2 \rightarrow 0; \\ k_{i+1}^2 \rightarrow 0; \\ (k_i k_{i+1})\rightarrow 0;
\nom}

So our gauge supercurrents are independent and nilpotent ones:

\disn{52}{
[k_i \tilde Y^{(i)}_n, k_i \tilde Y^{(i)}_m]=0; \cr [k_{i+1} \tilde Y^{(i+1)}_n, k_{i} \tilde Y^{(i)}_m]=0
\nom}
Let us notice that our choice for additional gauge conditions is appropriate for emission of $\pi$-mesons  (the case of usual quarks). It gives an explanation for massless $\pi$-mesons and correct amplitudes for $\pi$-mesons interaction \cite{Ku1}. But other quark flavours  bring us to gauge supercurrent constrains which contain not only fields with Lorentz indices $\widetilde Y^{(i)}$ but and
some part of  fields $\widetilde J^{(i)}$  for internal quantum numbers.

Now we are able to build spectrum generating algebra (SGA) for our set of states in the same manner as for the Neveu-Schwarz string model in \cite{Gr} by means of the operators of type of vertex operators (\ref{2}) of the conformal spin j to be equal to one.

We shall use the light-like vectors  $k^{(L)}_i$  from our vertices ($(k^{(L)}_i)^2=0$) and consider a state of the generalized momentum  $P^{(gen)}=p_0 + N k^{(L)}_i$.

\disn{53}{
\frac {p_0^2}2= -1; \\(k^{(L)}_i)^2=0; \\(k^{(L)}_i p_0)=1; \cr
(k^{(L)}_{i-1})^2=0; \\(k^{(L)}_ {i+1}   )^2=0;\cr
(k^{(L)}_ik^{(L)}_{i+1})\rightarrow 0; \\ (k^{(L)}_ik^{(L)}_{i-1})\rightarrow 0;
\nom}
 Transversal components of $k^{(L)}_i,p_0$ vanish $(p_0)_a=(k^{(L)}_i)_a=0$. The generalized mass of this state is given by :

\disn{54}{
\frac {M^{(gen)2}}2=\frac{(p_0+ Nk_i^{(L)})^2}2 =-1+N
\nom}

We define the transversal operators of SGA as corresponding vertex operators. All transversal SGA operators satisfy simple commutation algebra:

\disn{55}{
[(S^{(i)}_a)_n,(S^{(j)}_b)_m]=m\delta^{i,j}\delta_{a,b}\delta_{m+n,0}; \cr
[(S^{(i)}_a)_n,(B^{(j)}_b)_r]=0; \cr
\{(B^{(i)}_a)_r,(B^{(j)}_b)_s\}=\delta^{i,j}\delta_{a,b}\delta_{m+n,0}.
\nom}
 So we can construct similarly to the DDF states transversal states $|Phys\rangle$ from powers of the transversal SGA operators:

\disn{56}{
|Phys\rangle = \prod {((S^{(i)}_a)_{-n})^{\lambda (a,n)}}...|\Psi_0\rangle
\nom}
These states  satisfy all necessary gauge conditions.

Let us notice that all transversal SGA operators on the left side with $(i-1)$-th and $(i)$-th operators
($(i+1)$-operators vanish there)  can be defined with replacement of all $(i)$-fields to $(i-1)$-fields and vice versa of all $(i-1)$-fields to $(i)$-fields.  It is true and for all transversal SGA operators on the right side with $(i+1)$-th and $(i)$-th operators  ($(i-1)$-operators vanish there). They can be
defined with replacement of all $(i)$-fields to $(i+1)$-fields and vice versa of all $(i+1)$-fields to $(i)$-fields. This possibility to reformulate these sets of states allows to move from states of $i$-th section under consideration to states in $(i-1)$-th section and so on. Hence our consideration can be carried out up to ends of our amplitude (\ref{1}) and does not depend on the number $i$ of this section.

 Moving from these DDF type states to arbitrary states  we can obtain them as usually with help of ordered powers of the conformal generators $G_r, L_n$ and  of  powers  of  the  supercurrent operators $k^{(L)}_i \widetilde Y^{(L)(i)}$, $k^{(L)}_i \widetilde f^{(L)(i)}$ acting on $|Phys\rangle$  states.

After this taking into account the conditions (\ref{55}) we can repeat considerations in the Neveu-Schwarz model \cite{Gr} for  the  theorem about absence of ghosts in the spectrum of physical states in our case for the critical value of the number of effective dimensions.

In critical case the operators $G_{\frac {1}{2}}$ and

\disn{57}{
G_{\frac{3}{2}}+ 2 (G_{\frac {1}{2}})^3= G'_{\frac{3}{2}}
\nom}
define null states:

\disn{58}{
\{G'_{\frac{3}{2}},G_{\frac{-1}{2}}\}=0
\nom}

\disn{59}{
|S_{\frac{-3}{2}}\rangle = G'_{\frac{-3}{2}}|Phys\rangle; \\ \langle S_{\frac{3}{2}}|S_{\frac{-1}{2}}\rangle =0 \cr
| S_{\frac{-1}{2}}\rangle =G_{\frac{-1}{2}}|Phys\rangle; \\ \langle|S_{\frac{1}{2}}|S_{\frac{-1}{2}}\rangle =0
\nom}

\disn{60}{
\langle|S_{\frac{3}{2}} |S_{\frac{-3}{2}}\rangle =0
\nom}
The critical case corresponds to the condition (\ref{59}). It requires the condition (\ref{60}) to be satisfied:

\disn{61}{
L_0=\frac{1}{2}
\nom}
and  definite values of numbers of fields:

\disn{62}{
d_{crit}= 2d^{Lor} + 2d^{Int} =10
\nom}

For this case ($d^{Lor}=4$, $d^{Int}=1$)we can prove in the same way as in the Neveu-Schwarz model that if all constraints for physical states are fulfilled the norms of all physical states are non-negative. That means four  fields for all $Y$-fields i.e. $Y^{(i)}_{\mu},\mu=0,1,2,3$ and one component for  $J^{(i)}$-fields.

%%%%%%%%%%%%%%%%%%%%%%%%%
\section{Hadron vertex for u, d quark flavour.}
We formulate here the vertex $\hat {V_i}$ corresponding to the emission of ground state in an amplitude $A_N$ $N$ strings interaction. The simplest form for this composite string model has vertex operator $\hat V_i$ for $\pi$-meson emission:

\disn{2}{
\hat V_{i,i+1}(z_i)=z_i^{-L_0} \left[ G_r,\hat W_{i,i+1} \right] z_i^{L_0}, \\
\hat W_{i,i+1} = \hat R_i^{out}\hat R_{NS}\hat R_{i+1}^{in}
\nom}
The operators $\hat R_i^{out}$ and $R_{i+1}^{in}$ are defined by fields on $i$-th and $(i+1)$-th edging surfaces. The operator $\hat R_{NS}$ is defined by fields on the basic surface. They have the same structure as the operator

\disn{3}{
W_{NS}=:\exp ip_iX(1):=\exp(-p_i\sum_n\frac{a_{-n}}{n})\exp(-ip_iX_0)\exp(p_i\sum_n\frac{a_{n}}{n})
\nom}
of classical Neveu-Schwarz string model for both $Y$ and $J$ fields:

\disn{4}{
\hat R_i^{out}=\exp(\xi_i\sum_n\frac{J^{(i)}_{-n}}{n})\exp(k_i\sum_n\frac{Y^{(i)}_{-n}}{n})\exp(ik_i\bar Y_0^{(i)})\tilde\la^{(+)}_i\exp(-k_i\sum_n\frac{Y^{(i)}_{n}}{n})\exp(-\xi_i\sum_n\frac{J^{(i)}_{n}}{n})
\nom}

\disn{5}{
R_{i+1}^{in}=\exp(-\xi_{i+1}\sum_n\frac{J^{(i+1)}_{-n}}{n})\exp(-k_{i+1}\sum_n\frac{Y^{(i+1)}_{-n}}{n})\exp(-ik_{i+1}\bar Y_0^{(i+1)})\la^{(-)}_{i+1}\times \ns \times \exp(k_{i+1}\sum_n\frac{Y^{(i+1)}_{n}}{n})\exp(\xi_{i+1}\sum_n\frac{J^{(i+1)}_{n}}{n})
\nom}
where $\sum_ik_i=0$

\disn{6}{
\hat R_{NS}=\exp(-\dz_{i,i+1}\sum_n\frac{I_{-n}}{n})\exp(-p_{i,i+1}\sum_n\frac{a_{-n}}{n})\exp(-ip_{i,i+1}X_0)\times\ns\times\Gamma_{i,i+1}\exp(p_{i,i+1}\sum_n\frac{a_{n}}{n})\exp(\dz_{i,i+1}\sum_n\frac{I_{n}}{n})
\nom}

Here we have introduced $\la_{\al}$ operators to carry quark flavours and quark spin degrees of freedom. It must be stressed that $\la_i$ is an analog of $\exp(ik_iX_0)$ for field $\dd X$ and $\exp(ik_i\bar Y_0^{(i)})$ for field $Y^{(i)}$, since $\langle \la_i| J_0^{(i)}=\xi_i\langle \la_i| $ as $(\exp(ik_iX_0))\hat p_i=k_i(\exp(ik_iX_0))$ and $(\exp(ik_i\bar Y_0^{(i)}))Y_0^{(i)}=k_i(\exp(ik_i\bar Y_0^{(i)}))$ . This approach replaces usual transition to extra dimensions and allows introduce the quark quantum numbers in natural way. Besides this we obtain an attractive interpretation of the Chan-Paton factor in terms of two-dimensional fields.

\disn{7}{
\langle0|\tilde\la^{(+)}=0; \\ \la^{(-)}|0\rangle=0
\nom}

\disn{9}{
\left\{ \tilde\la^{(-)}_{\al},\la^{(+)}_{\be}\right\}=\de_{\al,\be},\\\tilde\la=\la T_0,\\T_0=\gamma_0\otimes\tau_2.
\nom}
Also were used values:

\disn{8}{
p_{i,i+1}=\hat\be_{in}^{(i+1)}k_{i+1}-\hat\be_{out}^{(i)}k_{i}, \\ \dz_{i,i+1}=\hat\al_{in}^{(i+1)}\xi_{i+1}-\hat\al_{out}^{(i)}\xi_{i}.
\nom}
Here $\xi$ is some universal matrix over quark flavours. So we give some relation between momenta (charges) flowing into the basic surface and into edging surfaces. Operators $\hat\al, \hat\be$ are defined by following properties:

\disn{35}{
\hat\be_{out}^{(i)}=\tilde\la^{(+)}_i\be_{out}\la^{(-)}_i, \\ \hat\al_{out}^{(i)}=\tilde\la^{(+)}_i\al_{out}\la^{(-)}_i, \cr [\be,\al]=0,\\\be^2=\al^2=1.
\nom}
However under consideration of tree diagrams it is possible to take both matrices to be equal to one without loss of generality. More careful consideration of these matrices $\hat\al, \hat\be$ has a quite definite sense for loop diagrams including nucleon states.

As it was mentioned above, we have to fulfil conditions for momenta: $k_i^2=k_{i+1}^2=0$, $k_ik_{i+1}=0$. These conditions lead to $m_{\pi}^2=0$.

%%%%%%%%%%%%%%%%%%%%
\section{Amplitude $\pi + \pi \rightarrow \pi + \pi$}
So we have $\pi$-meson emission vertex that coincides with the simplest vertex (\ref{2}). We require $\xi_i^2=\xi_{i+1}^2=1/2$ in order to have the conformal spin of this vertex to be equal to one.

Using vertex operator (\ref{2}) we can write $A_{\pi\pi}$ as the following integral:

\disn{67}{
A_{\pi\pi}=g^2\int_0^1 dz\; z^{-\frac3{2}}x^{(-\frac{p_{13}^2 }2+
\frac{\xi_{1}^2}2+\frac{\xi_{3}^2}2-\frac{k_{1}^2}2-\frac{k_{3}^2}2+\frac{\zeta_{13}^2}2)+\frac1{2}}(1-z)^{-p_{23}p_{34}+\zeta_{23}\zeta_{34}+k_{3}^2-\xi_{3}^2-1}\times \cr \times(-p_{23}p_{34}+\zeta_{23}\zeta_{34}+k_{3}^2-\xi_{3}^2)
\nom}
So we obtain this amplitude as a simple beta function:

\disn{68}{
A_4 = g^2 \frac{\Gamma (1- \alpha^t_0-\frac1{2}t)\Gamma (1-
\alpha^s_0-\frac1{2}s)} {\Gamma(1-\alpha^t_0-\frac1{2}t-\alpha^s_0-\frac1{2}s))} Tr(\Gamma_{12}\Gamma_{23}\Gamma_{34}\Gamma_{41})
\nom}
with $t=p_{13}^2$, $s=p_{34}^2$;  $\alpha^t_0=1-\frac{\xi_{1}^2}2-\frac{\xi_{3}^2}2+ \frac{k_{1}^2}2+ \frac{k_{3}^2}2-\frac{\zeta_{13}^2}2)$; $\alpha^s_0=-\frac{p_{23}^2}2+\frac{p_{34}^2}2-\zeta_{23} \zeta_{34}-k_{3}^2+\xi_{3}^2$. Now we take in to account $k_i^2=k_{i+1}^2=0$, $\xi_i^2=\xi_{i+1}^2=1/2$ and derive $\zeta_{13}=\xi_{1}-\xi_{3}=0$, $\zeta_{23}=\zeta_{34}=0$; $p_{23}^2=p_{34}^2=m_{\pi}^2=0$ and $\alpha^t_0=\alpha^s_0=\frac1{2}$.
So finally we have:

\disn{25}{
A_{\pi\pi}=-g^2Tr(\Gamma_{12}\Gamma_{23}\Gamma_{34}\Gamma_{41})\frac{\Gamma(1-\al_t^{\rho})\Gamma(1-\al_s^{\rho})}{\Gamma(1-\al_t^{\rho}-\al_s^{\rho})}
\nom}
with $\al_t^{\rho}=\frac{1}{2}+\frac{t}{2}$ and $\al_s^{\rho}=\frac{1}{2}+\frac{s}{2}$.

Now we are going to consider a partial expansion of (\ref{25}) for testing of absence of states with negative norm in spectrum of composite superstring model. It is easy to carry out this test for the case of scalar state to be second daughter state of the leading state with spin $J=2$. The partial expansion has the following form:

\disn{36}{
A_4=g^2\frac{r_2P_2(z)+r_1P_1(z)+r_0P_0(z)}{m_2^2-t},
\nom}
where $P_i(z)$ are Legendre polynomials. We need $r_0$ to be non-negative. For the case of $\al_t^{\rho}\rightarrow2$ we obtain the following expression for $r_0$:

\disn{37}{
r_0=\frac{3}{4}+\frac{x}{2}+\frac{x^2}{12}=\frac{1}{12}(x+3)^2,\\ x=4m_{\pi}^2-t
\nom}
 It is easy to see that $r_0$ is non-negative for any real $x$.

Now we can consider this expression for arbitrary $\al_0^{\rho}$. Then (\ref{37}) takes form:

\disn{64}{
r_0=\al_0^{\rho}(\al_0^{\rho}+1)+\frac{2\al_0^{\rho}+1}{4}x+\frac{1}{12}x^2, \\  x=4m_{\pi}^2-t
\nom}
The discriminant of this expression depends on $\al_0$:

\disn{65}{
D=-\frac{1}{12}(\al_0^{\rho}-\frac{1}{2})(\al_0^{\rho}+\frac{3}{2})
\nom}
As can be seen that for $\al_0^{\rho}\ge \frac{1}{2}$  there are no states with negative norm. (The case of $\al_0\le -\frac{3}{2}$ is excluded because the  asymptotic consideration of scalar daughter states for $J\gg 1$ leads to condition $\al_0^{\rho}>\frac{1}{3}$.) On the other hand we have condition $\al_0^{\pi}\le 0$ to avoid tachyon on $\pi$-trajectory. We require Adler-Weinberg condition to be fulfilled for $A_{\pi \pi}$. It coincides with   vanishing of gamma function in denominator in (\ref{25}):

\disn{33}{
1-2\al_0^{\rho}-2\al'm^2_{\pi}=1-2\al_0^{\rho}+2\al_0^{\pi}
\nom}
it leads to
\disn{34}{
\al_0^{\pi}=\al_0^{\rho}-\frac{1}{2}
\nom}
 To satisfy this condition we have to take $\al_0^{\rho}$ to be less or equal to $\frac{1}{2}$. So we have $\al_0^{\rho}=\frac{1}{2}$.

%%%%%%%%%%%%%%%%%%%%%%%%%
\section{Hadron vertex for u,d,s quark flavours}
We cannot build a $K$-meson emission vertex in the same Neveu-Schwarz form as for $\pi$-meson emission $(k_i^2=k_{i+1}^2=0)$. So far as then we shall lose supercurrent conditions for the $s$-quark edging surface with $k_{i+1}^2 \ne0)$ and hence we lose non-negativity of physical states norms. So we have to construct another possible form, a vertex by Bardakci-Halpern type for $K$-meson emission \cite{Ku2}.

So we have to construct another form of $\hat V_{i,i+1}$ in the case of $K$-meson:

\disn{10}{
\hat V_{i,i+1}(z_i)=z_i^{-L_0} \left\{ G_r,\tilde W_{i,i+1} \right\} z_i^{L_0}, \\ \tilde W_{i,i+1}=\hat F\hat W_{i,i+1}, \\\hat W_{i,i+1} = \hat R_i^{out}\hat R_{NS}\hat R_{i+1}^{in}
\nom}
where $\hat R_i^{out}$, $\hat R_{i+1}^{in}$, $\hat R_{NS}$ are (\ref{4}), (\ref{5}), (\ref{6}) respectively. Here we have $i$-th edging surface for usual (u, d) quark flavours with $k_i^2=0$, $k_ik_{i+1}=0$, $\xi_i^2=\frac{1}{2}$ and the $(i+1)$-th edging surface for s-quark. The conformal spin of the vertex operator $\hat V_{i,i+1}$ ought to be equal to one. The conformal spin of operator $\hat F$ is equal to $\frac{1}{2}$ with respect to Bardakci-Halpern form. So we require that  operator $\hat W_{i,i+1}$ to have conformal spin equal zero. Thus operator $\hat F$ must be composed of fields with conformal spin equal to one half. These fields are superpartners of the fields propagating on basic and edging surfaces. Its structure will be discussed below.

There are two orthogonal light-like supercurrent conditions: the old one

\disn{11}{
k_i\tilde f^{(i)}=k_i(f^{(i)}+\hat \be_iH), \\  k_i\tilde Y^{(i)}=k_i(Y^{(i)}+\hat \be_i\dd X)
\nom}
and the new second one

\disn{12}{
\tilde f_s=k_{i+1}( f^{(i+1)}+\hat \be_{i+1}H)-\xi_i \Phi^{(i)}+\xi_{i+1}\Phi^{(i+1)}+(\xi_{i+1}\hat\al_{i+1}-\xi_i\hat\al_i)\Theta,  \cr
\tilde Y_s=-k_{i+1}(Y^{(i+1)}+\hat\be_{i+1}\dd X)-\xi_i J^{(i)}+\xi_{i+1} J^{(i+1)}+(\xi_{i+1}\hat\al_{i+1}-\xi_i\hat\al_i)I
\nom}
We require

\disn{13}{
-k_{i+1}^2+\xi_i^2+\xi_{i+1}^2-\xi_i\xi_{i+1}=0
\nom}
and $k_i^2=0$ in order to have the conformal spin of operator $\hat V_{i,i+1}$ to be equal to one (here we are taking to account the conformal spin of $\hat F$ equal to $\frac{1}{2}$) and the light-likeness of (\ref{11}), (\ref{12}) simultaneously. New vertex operator (\ref{10}) satisfies these supercurrent conditions.

Now we will discuss the structure of the operator $\hat F$. It must satisfy a set of conditions. First of all it ought to be orthogonal to the combination of fields composing the exponent of $\hat W_{i,i+1}$ in order to avoid infinity which can appear after normal ordering of vertex operator $\hat V_{i,i+1}$. Furthermore operator $\hat F$ has to satisfy supercurrent conditions (\ref{11}), (\ref{12}). In our case these two requirements coincide. There are three possible independent forms of such operator. However only one form does not lead to ancestors. It contains only fields propagating on the $(i+1)$-th edging surface. So we have:

\disn{14}{
\hat F= \tilde \gamma (k_{i+1}Y^{i+1})+\frac {\tilde \gamma m_K^2}{\xi_{i+1}}J^{i+1}
\nom}

The equation (\ref{13}) gives us value range of $k_{i+1}^2=k_{s}^2= m_K^2$. It is interesting that $k_{s}^2$ has its minimal value corresponding to $m_K \approx 475MeV$. However we have to increase the value of $k_{s}^2$ to avoid tachyon on the $\varphi$-meson trajectory. As it would be seen below minimal appropriate values of the momenta are:

\disn{15}{
\xi_{i+1}^2 \equiv \xi_s^2=k_s^2=\frac{1}{2}
\nom}

%%%%%%%%%%%%%%%%%%%%%%%%%%
\section{Amplitude $\pi + K \rightarrow \pi + K$}
To calculate amplitude $A_4(\pi + K \rightarrow \pi + K)$ we have to use vertices of both types (\ref{2}) and (\ref{10}). So we can write the following integral:

\disn{69}{
A_{\pi K}=g^2\la_K^2Tr(\Gamma_{12}\Gamma_{23}\Gamma_{34}\Gamma_{41})\int_0^1 dz\; z^{-\frac{m_K^2}{2}-\frac{t}{2}+\frac{\xi^2}{2}+\frac{\xi_s^2}{2}+\frac{(\xi_s-\xi)^2}{2}-\frac{3}{2}} (1-z)^{-\xi^2-\frac{s}{2}} \times \cr \times \frac{z}{(1-z)}(-\de_s)\frac{m_K^2}{\xi_s^2}\tilde\gamma^4 (-\xi^2-\frac{s}{2})
\nom}
where $\de_s$ is

\disn{17}{
\de_s=\xi_s^2-k_s^2.
\nom}
Taking in to account $\xi^2 \equiv \xi^2_i=1/2$ and (\ref{13}), we obtain $A_{\pi K}$ as a simple beta function:

\disn{16}{
A_{\pi K}=-g^2 Tr(\Gamma_{12}\Gamma_{23}\Gamma_{34}\Gamma_{41})\la_K^2\de_s\frac{m_K^2}{\xi_s^2}\tilde\gamma^4 \frac{\Gamma(1-\al^{K^*}_t)\Gamma(1-\al^{\rho}_s)}{\Gamma(1-\al^{K^*}_t-\al^{\rho}_s)}
\nom}
with $\al^{K^*}_t=-\frac{m_K^2}{2}+\frac{1}{2}+\frac{t}{2}$ and $\al^{\rho}_s=\frac{1}{2}+\frac{s}{2}$.

The analysis of a partial expansion of (\ref{16}) for $\al^{K^*}_t\rightarrow2$ leads us to the following expression for $r_0$:

\disn{38}{
r_0=\frac{3}{4}+\frac{R}{2}+\frac{R^2}{12}=\frac{1}{12}(R+3)^2,\\ R=2(m_{\pi}^2+m_{K}^2)-\frac{(m_{\pi}^2-m_{K}^2)^2}{t}-t
\nom}
 Again it is easy to see that $r_0$ is non-negative for any real $R$.

%%%%%%%%%%%%%%%%%%%%%%%%%%
\section{Amplitude $K+K\rightarrow K+K$}
To calculate the amplitude $A_4(K + K \rightarrow K + K)$ one should use the vertex of the type (\ref{10}) only. This expression could be written in the following form:

\disn{39}{
A_{K K}=g^2Tr(\Gamma_{12}\Gamma_{23}\Gamma_{34}\Gamma_{41})\la_K^4\de_s^2\frac{m_K^4}{\xi_s^4}\tilde\gamma^4 \ls-\frac{\Gamma(1-\al^{\varphi}_t)\Gamma(1-\al^{\rho}_s)}{\Gamma(1-\al^{\varphi}_t-\al^{\rho}_s)}+\frac{\Gamma(-\al^{\varphi}_t)\Gamma(1-\al^{\rho}_s)}{\Gamma(1-\al^{\varphi}_t-\al^{\rho}_s)}\rs
\nom}
with $\al^{\rho}_s=\frac{1}{2}+\frac{s}{2}$ and $\al^{\varphi}_t=\frac{t}{2}-\de_s$ where parameter $\de_s$ is (\ref{17}).

\begin{figure}[h!]
\center{\includegraphics[width = 8 cm]{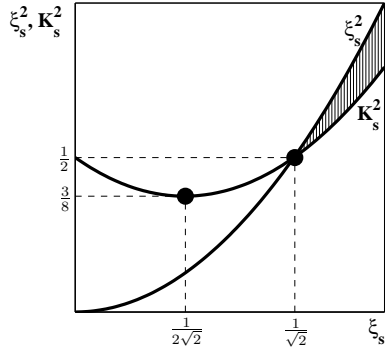}
\caption{Connection between  $\xi_s^2$ and $k_s^2$ for $\de_s>0$
(hatched area).} \label{Grafik}}
\end{figure}

As can be seen above the value of $\de_s$ have an influence on the intercept of $\varphi$-meson trajectory. If we have $\de_s<0$, hence there is a tachyon. So we have to find a relation between values of $\xi_s^2$ and $k_s^2$. We can consider (\ref{13}) and derive following expression:

\disn{18}{
k_s^2=(\xi_s-\frac{1}{2}\xi)^2+\frac{3}{8}
\nom}
Comparison of $\xi_s^2$ and $k_s^2$ (see Fig.~(\ref{Grafik})) leads us to the following conditions:

\disn{66}{
\xi_s^2\ge\frac{1}{2}, \\  \al' m^2_K=\frac{1}{2}m^2_K\ge \frac{1}{4}
\nom}
Similar consideration is literally allowably for more massive quark flavours ($c$, $b$, $t$) with corresponding values for $\xi_c$, $\xi_b$, $\xi_t$, $ m^2_c$, $ m^2_b$, $ m^2_t$.

Consideration of the partial expansion of (\ref{39})  for $\al^{\varphi}_t\rightarrow2$ leads us again to the following expression for $r_0$:

\disn{40}{
r_0=\frac{3}{4}+\frac{y}{2}+\frac{y^2}{12}=\frac{1}{12}(y+3)^2,\\ y=4m_K^2-t
\nom}
And $r_0$ is non-negative for any real $y$.

Now we consider partial expansion for the cross-channel of (\ref{39}).  So for $\al^{\rho}_s\rightarrow2$ we have $r_0$ in the following form:

\disn{63}{
r_0=(1-\de_s)^2-\frac{(1-\de_s)}{2}z+\frac{1}{12}z^2, \\ z=s-4m_K^2
\nom}
And $r_0$ is positive for any real $z$ dew to negative discriminant $D=-\frac{1}{12}(1-\de_s)^2$.

%%%%%%%%%%%%%%%%%%%%%%
\section{Conclusion}
We have obtained simple Born amplitudes of $\pi$ and $K$ mesons in the composite superstring model. Next step is to find one-loop amplitudes for this approach and Born diagrams for nucleons and other baryons  interaction.

The authors wish to thank participants of Theoretical Division of PNPI seminars for discussions of this work.

\end{document}